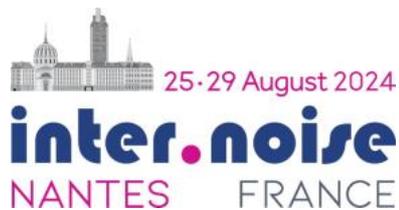

# Artificial intelligence in creating, representing or expressing an immersive soundscape

Rima Ayoubi[1]
Nantes Université, ENSA Nantes, École Centrale Nantes, CNRS, AAU-CRENAU, UMR 1563, F-44000 Nantes, France

Sang Bum Park[2]
School of Architecture and Engineering Technology
Florida Agricultural and Mechanical University
1983 South Martin Luther King, Jr., Blvd., Tallahassee, FL 32307

Laurent Lescop[3]
Nantes Université, ENSA Nantes, École Centrale Nantes, CNRS, AAU-CRENAU, UMR 1563, F-44000 Nantes, France

**ABSTRACT**

*In today's tech-driven world, significant advancements in artificial intelligence and virtual reality have emerged. These developments drive research into exploring their intersection in the realm of soundscape. Not only do these technologies raise questions about how they will revolutionize the way we design and create soundscapes, but they also draw significant inquiries into their impact on human perception, understanding, and expression of auditory environments. This paper aims to review and discuss the latest applications of artificial intelligence in this domain. It explores how artificial intelligence can be utilized to create a virtual reality immersive soundscape, exploiting its ability to recognize complex patterns in various forms of data. This includes translating between different modalities such as text, sounds, and animations as well as predicting and generating data across these domains. It addresses questions surrounding artificial intelligence's capacity to predict, detect, and comprehend soundscape data, ultimately aiming to bridge the gap between sound and other forms of human-readable data.*

## 1. INTRODUCTION

Artificial intelligence (AI) technologies initially witnessed successful advancements during the 1950s and 1960s [1], particularly with programs such as the General Problem Solver (GPS) [2]. However, it wasn't until 2022, with the release of ChatGPT-3.5 by OpenAI, that generative AI became widely accessible [3], sparking considerable interest among researchers, industry professionals and individuals from various fields including soundscape, art and architecture [6]. This release facilitated unprecedented accessibility to generative AI technology, opening up opportunities for research, experimentation and innovation. Consequently, we are witnessing today an accelerated development and deployment of AI technologies across various sectors. This is leading to a significant increase in the number and diversity of AI applications which are impacting professional practices and introducing new methodologies and innovative tools to streamline processes.

---

[1] rima.ayoubi@crenau.archi.fr

[2] sang.park@famu.edu

[3] laurent.lescop@nantes.archi

AI, defined as the system's adeptness in interpreting external data, learning from it, and using these insights to achieve specific goals [4], incorporates various technologies. Among these, Generative AI emerges as a significant field within AI, focusing on generating or predicting new data from existing datasets [7]. Powered by deep learning, a subset of machine learning grounded in artificial neural networks, Generative AI surpasses traditional AI methods by actively seeking out data to address user queries or tasks [6]. Deep learning techniques are employed to undergo extensive training with vast datasets, aiming to generate human-like new content such as images, text, audios, videos and 3d models [1]. The diverse capabilities of AI-generated content (AIGC) offer a wide range of applications. Through AI, data can seamlessly transition from one form to another, such as text to image, the combination of image and text to video, or sketch and text to 3D model [1].

This accelerated development and deployment of AI technologies finds direct application in online platforms and through software, including 3D modeling and game engine software such as Unity and Blender, facilitated by APIs. Application Programming Interfaces (APIs) enable users to integrate generative and interactive AI into their systems or applications [3]. In addition to API's, many generative AI tools are offered as open source, eliminating cost barriers for usage and making generative AI more approachable and interactive [3]. With advancing AI capabilities, users can engage in diverse and dynamic interactions, as is the case with conversational interfaces such as ChatGPT and Convai [1].

AI is presenting both challenges and opportunities across diverse fields including the domain of soundscape. With the emergence of generative and interactive AI technology, areas and tasks that were considered complex or time-consuming are entering a new phase. This technology enables the creation of new applications that were previously considered impractical or unattainable for automation. Indeed, generative AI is currently revolutionizing our approach to work, leading to the development of new ideas, and ultimately to the emergence of completely new models of work [7].

In the gaming industry, users have already integrated AI in real-time 3D scene rendering, enhancing both visual and sound content generation [1]. Furthermore, the implementation of AI-embedded technologies into virtual reality enhances various aspects such as content generation and user interaction, particularly with speech and sound recognition [8]. As the case with the recent integration of AI into the Unity game engine, aimed at optimizing asset and script creation, enhancing problem solving capabilities and facilitating VR interaction development. Examples such as Wit.ai, Convai, Unity Muse, and Zibra AI demonstrate the diverse applications of AI technology within Unity environments.

Due to the complexity of soundscape and its design, numerous researchers have started using virtual reality as a tool to construct immersive three-dimensional auditory and visual environments [9-13]. Moreover, users' interpretations of a specific soundscape can vary widely, due to their own experiences, emotions, sensibilities and cognitive abilities [10-14]. To enhance soundscape design, artificial intelligence offers a promising solution by using deep learning from extensive datasets to provide relevant recommendations [11].

Therefore, we need to examine and contemplate in this article how we can use and develop generative and interactive AI as a new tool for soundscape creation. This paper explores key applications of generative and interactive AI in generating, representing or expressing an immersive soundscape. By offering a new perspective on this process, it analyzes new opportunities and discusses the challenges they present.

## 2. METHOD

In this research, and as a part of the "FAMU Digital Documentation Project" [12], we undertook a study involving the First Presbyterian Church of Tallahassee in Florida. The goal is to develop an immersive and interactive soundscape experience within the church environment using AI applications. To achieve this, we utilized several visual elements, including:

- A point cloud representing the interior of the First Presbyterian Church of Tallahassee, generated from a 360-degree video using Metashape Agisoft.
- A point cloud of people, generated with luma AI, a text-to-3D-model tool.
- Additionally, Blender, an open-source 3D computer graphics software tool, was used to decimate and export the visual data into a PLY file format for later integration into Unity engine.

Our approach involved proposing a systematic process, starting with AI-driven sound generation, leading to the development of an immersive soundscape within a virtual reality environment, and its subsequent visualization. The method consists of going through the following steps:

Step 1: Generate sounds with AI applications.
- Soundscape composition: This step involves describing the composition of a church soundscape, listing all the sounds typically found in such an environment.
- AI applications and results: In this step, three different AI applications are utilized to generate the sounds listed in the previous section. Each application may employ different techniques, algorithms, or datasets to produce sound outputs. The results generated by each AI application are compared and evaluated for their realism, fidelity, and suitability for creating a convincing church soundscape. This comparison helps determine which AI tools are most effective for generating the desired auditory elements.

Step 2: Create an immersive virtual soundscape inside Unity engine.
- 3D audio Spatialization: The sounds generated in the previous step are integrated into a virtual environment within the Unity engine. This includes spatializing the audio to simulate realistic 3D sound perception for the user. By assigning spatial coordinates to each sound source, the audio can be perceived as coming from different directions and distances within the virtual space.
- Interaction system with wit.ai: An AI application that enables interaction within the virtual environment through natural language processing. Users will be able to control or influence aspects of the soundscape using voice commands processed by wit.ai, adding an interactive element to the experience.

Step 3: Visualize the soundscape in Virtual Reality using visual effects (VFX) in Unity engine.

In this phase, we used visual effects (VFX) within the Unity Engine to merge the auditory and visual dimensions, assisted by ChatGPT. Through dynamic particle effects, we translated the audio clip into a visual representation, animating the particles' scale and color in response to the audio frequencies. Our goal is to explore AI in the interpretations of soundscapes.

## 3. RESULTS AND ANALYSIS

Throughout this section, we will examine each step of the workflow process needed to create the immersive and interactive soundscape of the Church. At each stage, we will present the AI applications and analyze the resultant outcomes.

### 3.1. Generate sounds with AI applications

1) Soundscape composition

A soundscape is a multi-layered construct of sounds with different spectral and temporal properties, which may not necessarily emerge at the same level [5]. Therefore, to create the church's soundscape, we will generate each sound component separately and then place them within the 3D model. This method enables precise control over each sound element, especially for interactive purposes within the immersive experience.

The auditory environment within a church is composed of a spectrum of sounds, collectively shaping its distinct ambiance. In this study, we aim to identify the predominant auditory elements commonly encountered within a church:
- Sound of organ music
- Sound of a choir singing
- Sound of creaking wooden pews
- Sound of footsteps
- Sound of people whispering
- Sound associated with the opening and closing of doors

2) AI applications and results

To produce the specified sounds, we chose three AI applications for their free and easy-to-use accessibility. However, it is important to note that generative AI technology is rapidly evolving, with platforms and applications advancing and changing quickly due to their usage and the datasets they employ. Here is a list of the AI applications we used in this task, along with the subsequent outcomes of the generated audio:
- Stable Audio, an AI-generated audio tool that operates on text-to-audio and audio-to-audio generation. It has been trained using a licensed dataset sourced from the AudioSparx music library.
- Audiogen, an autoregressive generative model that produces audio samples based on text inputs. It utilizes a specific, pre-learned method for representing audio data. This representation is derived from previous training or learning processes [15].
- OptimizerAI develops AI technology to generate audio resources needed for game and content development. It can produce simple sound effects containing a single action or sound. As it was recently released, the model is ongoing updates and training to enhance its capabilities over time. With the recent v0.2 update, they have improved the model's performance on UI sounds.

Prompt: *generate the sound of an organ music inside a church*
- Stable Audio: generates a single option for 45 seconds, but it does not accurately replicate the sound of an organ within a church with significant background noise.
- Audiogen: produces four distinct options, each lasting 10 seconds, and effectively replicates the sound of an organ inside a church.
- OptimizerAI: generates five different options, each lasting 3 seconds, and effectively replicates the sound of an organ inside a church.

Prompt: *generate the sound of a choir singing inside a church*
- Stable Audio: effectively replicates the sound.
- Audiogen: replicates effectively the sound.
- Optimizer AI: replicates effectively the sound.

Prompt: *generate the sound of creaking wooden pews inside a church*
- Stable Audio: generates music rather than sound effects, with machine noise present in the background.
- Audiogen: The 4 options do not resemble the sound of creaking wooden pews; instead, they feature a metallic effect accompanied by some beats.
- Optimizer AI: replicates effectively the sound.

Prompt: *generate the sound of footsteps inside a church*
- Stable Audio: does not accurately replicate the sound with significant background noise. We attempted a variety of prompts. For *slow footsteps,* it generates a sound that matches but with an outdoor ambiance. When using the prompt *slow footsteps echoing inside*, the output does not align well, resulting in a very noisy sound.
- Audiogen: the output does not align well, resulting in a very noisy sound.
- Optimizer AI: replicates effectively the sound.

Prompt: *generate the sound of people whispering inside a church*
- Stable Audio: does not replicate the sound with significant background noise.
- Audiogen: generates a musical whispering sound inside the church.
- Optimizer AI: replicates effectively the sound.

Prompt: *generate the sound of opening and closing the doors inside a church*
- Stable Audio: generates a musical output that doesn't respond to the prompt.
- Audiogen: effectively replicates the sound after adding in the negative prompt option *echo and reverberation* to reduce the background noise.
- Optimizer AI: replicates effectively the sound.

Table 1: Comparison of Key Features and Limitations of AI Applications.

| AI APPLICATIONS | LIMITATIONS | KEY FEATURES |
|---|---|---|
| Stable Audio | - Free version:<br>  o 20 attempts / month<br>  o Track duration 45s<br>- Paid version 11.99$/month:<br>  o 500 attempts / month<br>  o Track duration 90s | - Stable Audio is trained on a dataset sourced from the AudioSparx music library, which makes it incompatible with sound effects generation. |
| Audiogen | - Free version:<br>  o 100 attempts / month<br>  o Track duration 10s<br>- Paid version 5$/month:<br>  o 1000 attempts /month<br>  o Track duration 90s | - Sign in with discord<br>- Generates sounds that correspond to the prompt, although occasional background noises may be present.<br>- It does not produce a single sound output, often requiring multiple attempts to refine the output. |
| OptimizerAI | - Unlimited free usage (beginning of this server's journey) | - Sign in with discord<br>- Generate simple sound effects containing a single action or sound<br>- Works well with sound effects |

## 3.2. Create an immersive virtual soundscape inside Unity engine
1) 3D audio Spatialization

In this phase of the research, the Unity engine was employed to develop a virtual soundscape augmented with 3D audio spatialization. Each auditory element was placed within the virtual church environment, aligning with precise locations on the map. For instance, the placement of organ music is configured to originate proximate to the organ, while whispers are situated in close proximity to the point cloud of people within the virtual space (see Figure 1).

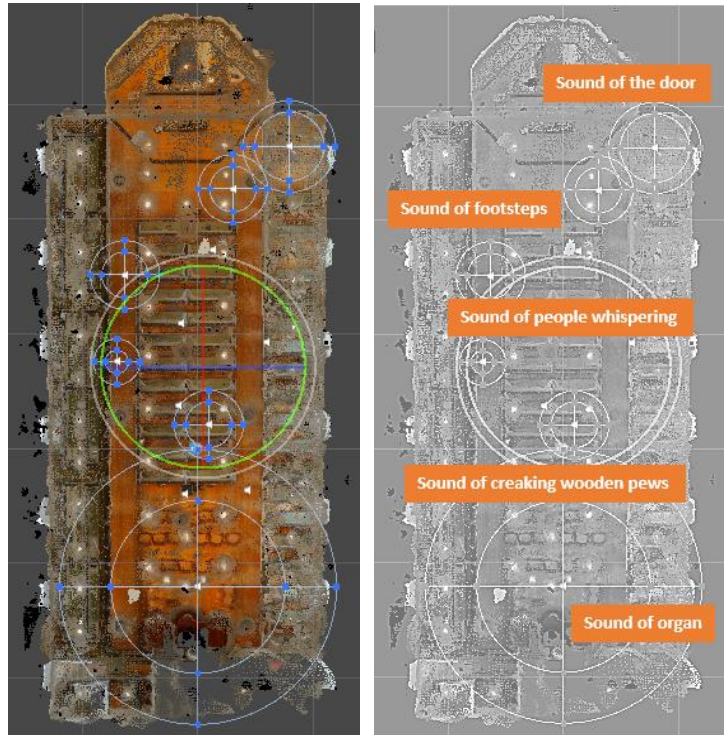

Figure 1: Spatialized Sound Mapping: Church Layout with Audio Representation.

Within Unity, we employed the Oculus Spatializer to enhance the immersive quality of the soundscape (see Figure 2). Our focus primarily revolved around optimizing audio sources, particularly through the manipulation of spatial blend parameters, to heighten the sense of immersion (see Figure 3). Additionally, we implemented an audio reverb zone to customize the acoustic characteristics of the environment to our needs. This feature allows us to define specific reverberation properties, further enhancing the realism of the auditory experience within the virtual church setting (see Figure 4).

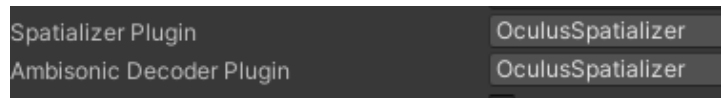

Figure 2: Oculus Spatializer Plugin within Unity.

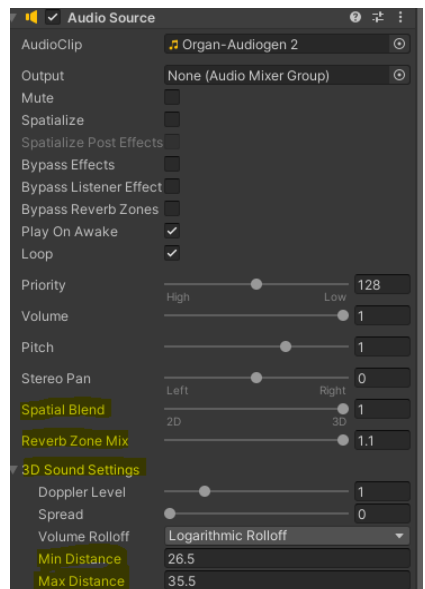

Figure 3: Audio source parameters within Unity.

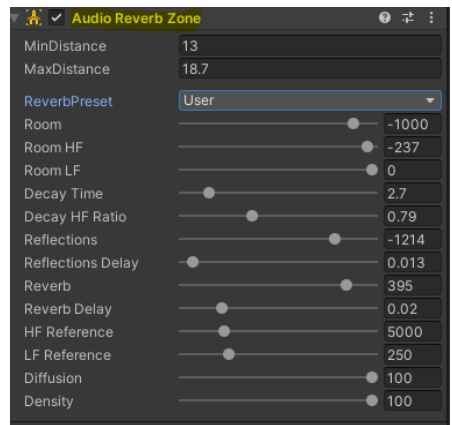

Figure 4: Audio Reverb Zone properties within Unity.

As we choose to loop the audio continuously to accompany users as they navigate the virtual space with their headsets, we'll need to implement a mechanism to stop the audio playback once the user moves beyond a certain distance. To achieve this functionality, we used ChatGPT to assist in scripting the necessary logic (see Figure 5 and 6).

```
using UnityEngine;

public class AudioDistanceController : MonoBehaviour
{
    public AudioSource audioSource; // The AudioSource component
    public float maxDistance = 10f; // Maximum distance for full volume

    private void Update()
    {
        // Calculate the distance between the main camera and the audio source
        float distance = Vector3.Distance(Camera.main.transform.position, transform.

        // Calculate the volume based on the distance
        float volume = Mathf.Clamp01(1f - (distance / maxDistance)); // Volume decre

        // Set the volume of the audio source
        audioSource.volume = volume;
    }
}
```

Figure 5: Script Generated by ChatGPT: Final Output

Prompt: *write a unity script for an audio source to adjust its volume. When the main camera moves beyond the specified maximum distance, the volume of the audio source decreases to zero.*

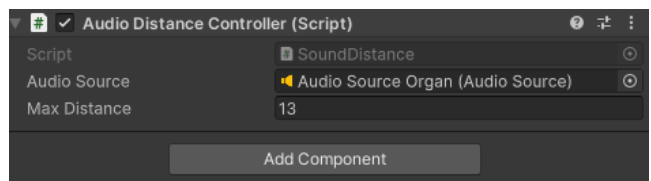

Figure 6: Script Output Integrated with Audio Source in Unity: Generated by ChatGPT

2) Interaction system with Wit.ai

In this segment of the study, our objective was to enable real-time interaction for a Natural User Interface (NUI) and facilitate user engagement with the 3D virtual soundscape. To accomplish this, we utilized the Meta XR Unity Package Manager (UPM) from the asset store - a free package that combines the Voice SDK and Wit.ai, thus enabling voice interactions. By using Wit.ai, we could train applications to recognize voice commands and trigger corresponding sound effects through speech recognition.

Several steps were involved in achieving this task:
- Training the machine via the Wit.ai to comprehend the intended interaction. This process entailed defining entities, assigning synonyms, and establishing roles and outcomes for each entity. Through training, the application learned to recognize specific words and their associated effects (see Figure 7).
- In Unity, configuring the setup after integrating the API and adjusting various settings to link the three systems: Wit.ai, Unity engine, and Oculus XR Rig (see Figure 8).

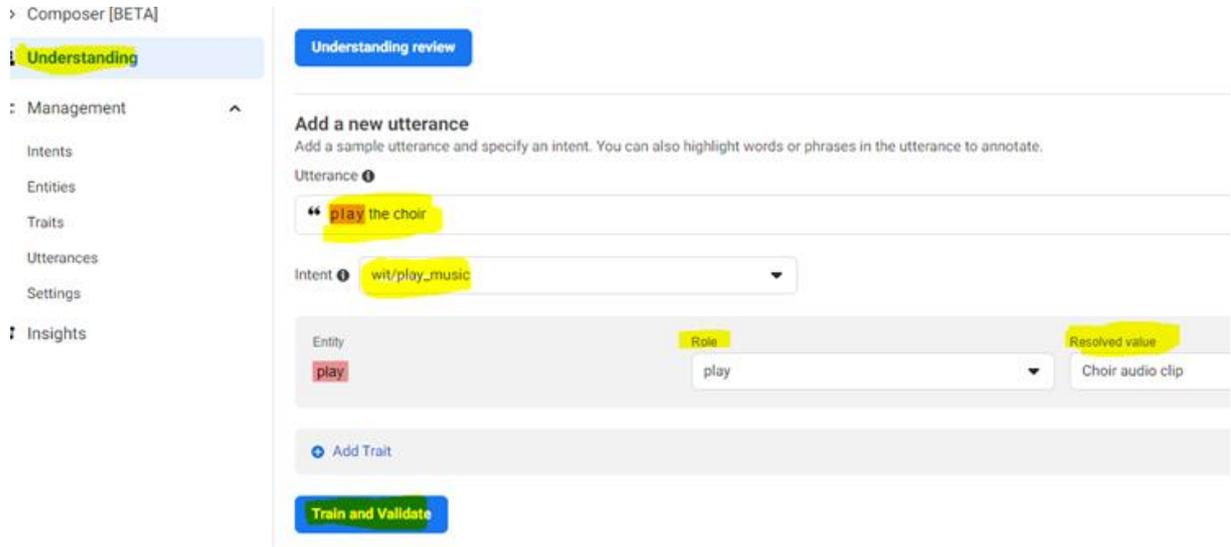

Figure 7: Wit.ai learning and training system.

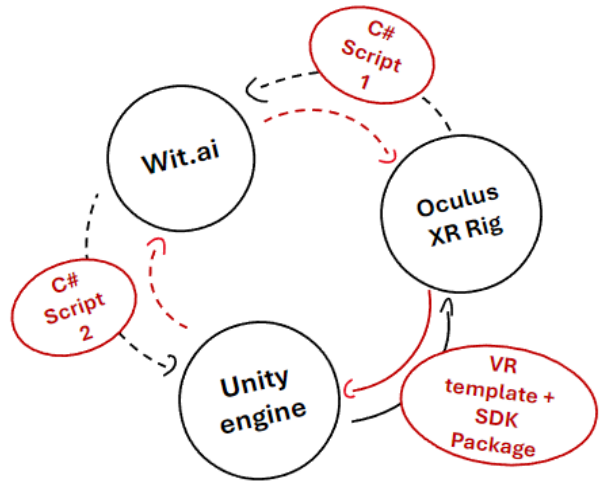

Figure 8: Wit.ai configuration logic with Oculus and Unity

For this setup, we used ChatGPT to generate two scripts that establish connectivity between the three systems: Wit.ai, the Unity engine, and the Oculus XR Rig.
- Script 1: To be attached to the Oculus Voice SDK's App voice experience, facilitates the connection between the Oculus XR Rig and Wit.ai by utilizing the TriggerPressed event.
- Script 2: to be linked to the Response Handler from Wit.ai, establishes communication with the Unity system to execute effects created in the scene. When a command is triggered by Wit.ai in response to the TriggerPressed event from Oculus, Script 2 prompts Unity to execute the specified action, such as playing music.

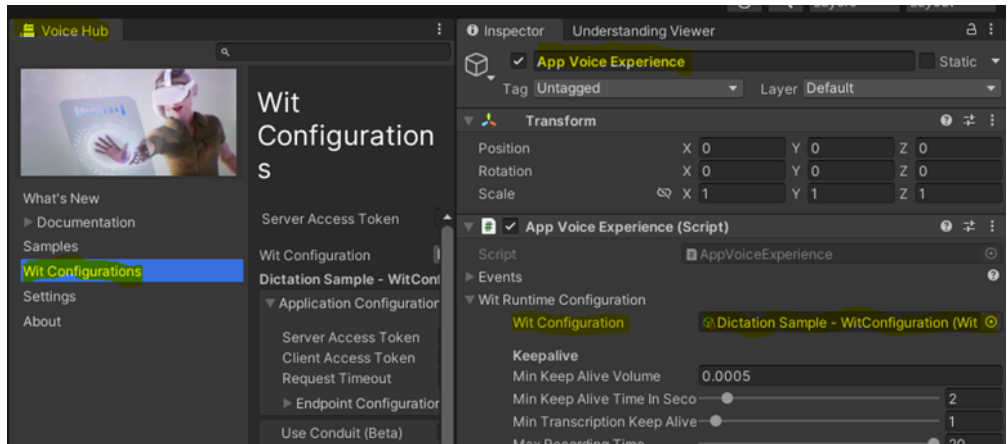
Figure 9: Wit.ai configuration in Unity.

### 3.3. Visualize the soundscape in Virtual Reality using visual effects (VFX) in Unity engine

To explore AI interpretations of soundscapes, we worked on generating C# script with the assistance of ChatGPT. The script is used to convert audio into samples containing spectrum data frequencies. Subsequently, each particle is attached to a sample, animating based on the frequency in the spectrum. This process is achieved using visual effects (VFX) in Unity (see Figure 10).

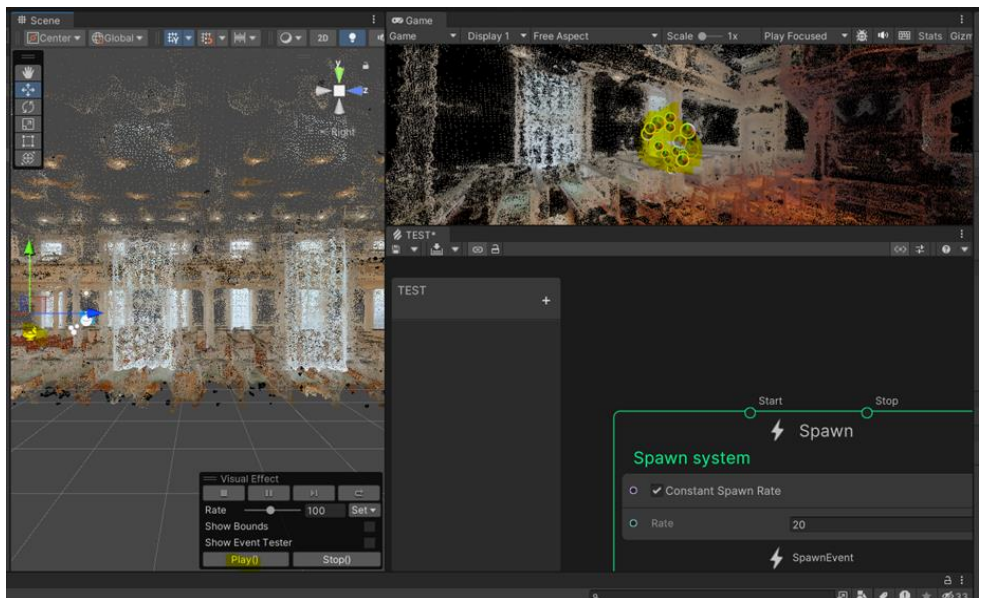
Figure 10: Adding visual effects (VFX) to the point cloud in Unity.

The script, to be linked to the visual effect graph created in the asset, is capable of:
− Extracting audio spectrum data to visualize an audio clip. With the use of the *GetComponents<AudioSource>()* and *GetSpectrumData(samples, channel, FFTWindow)* functions to retrieve the spectrum data from the audio source. Additionally, we specified the number of samples to be generated with the script.

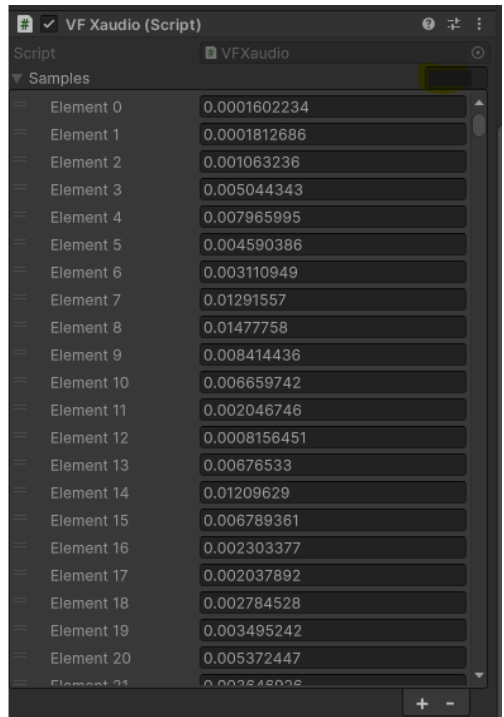

Figure 11: Script 1 generating samples with different spectrum data while the audio is playing.

- Linking each spectrum with the prefab we've established in the scene using the Visual Effect Graph (particles, cubes, point clouds). This enables synchronization of the prefab's particles (including scale and color) with the audio playback through VFX properties.

## 4. DISCUSSION AND CONCLUSIONS

From the initial phase of the study, it was observed that AI applications for generating sound effects are still in their early stages. While OptimizerAI provides a glimpse into this potential, a new release from ElevenLabs, AI Sound Effects for OpenAI Sora, has garnered significant anticipation with over 11,000 people on its waitlist. Current advancements, exemplified by OptimizerAI, demonstrate the feasibility of generating simple sound effects containing a single action or sound, thereby opening the way for new methodologies in soundscape creation. As AI applications rely on continuous learning and adaptation from datasets, this ongoing process is crucial for refining and enhancing their capabilities as they progress. Especially that the gaming industry serves as a driving force for advancing these applications, causing considerations about the urgency of adapting to technological advancements to effectively train AI to meet the demands of scientific research.

In the second phase of the study, we explored the utility of AI as an assistant in problem-solving and script generation, exemplified by ChatGPT's contribution to achieving our objectives. Additionally, we investigated AI's role in augmenting human interaction within virtual environments, which drive research into studying its influence on human perception, and comprehension of auditory environments.

In the third phase of the study, we examined how AI can interpret sounds through visualization. However, the lack of explainability and transparency in audio visualization raises several concerns. Users may struggle to interpret and comprehend the output, limiting their ability to identify potential errors and affecting their trust in the system.

Within the realm of Audio-Visual interaction, AI-Driven Adaptive Solutions are integrated into game engines like Zibra AI for Unity to meet the gaming industry's evolving needs for advanced visual effects (VFX) solutions and enhanced rendering capabilities. These advancements show potential for enriching audio-visual interactions and interpretations by facilitating real-time rendering channels. As AI continues to evolve, addressing issues of transparency and user trust remains important to ensure the effectiveness and acceptance of these technologies.